# Single-pair charge-2 Weyl–Dirac composite semimetals

Hui-Jing Zheng[*], Ke-Xin Pang[*], Yun-Yun Bai, Yanfeng Ge, and Yan Gao[†]

*State Key Laboratory of Metastable Materials Science and Technology & Hebei Key Laboratory of Microstructural Material Physics, School of Science, Yanshan University, Qinhuangdao 066004, China*

The Nielsen–Ninomiya theorem requires that the total topological chiral charges in a crystal vanish, a constraint typically satisfied by identical nodes like Weyl–Weyl pairs. Whether a minimal heterogeneous configuration—comprising a single Weyl point (WP) and a single Dirac point (DP)—can exist in an electronic system has remained unresolved. Here, by systematically classifying all 1651 magnetic space groups (MSGs), we reveal that only 14 MSGs without spin-orbit coupling (SOC) and 10 MSGs with SOC are compatible with this exotic state. Furthermore, for nonmagnetic crystals, this configuration is uniquely realized in the spinless limit of chiral space groups 92 and 96. Guided by this principle, we predict an ideal realization in chiral three-dimensional boron allotropes (SDHBN-$B_{28}$ enantiomers). First-principles calculations unveil a $|C| = 2$ WP at the $\Gamma$ point and a $|C| = 2$ DP at the A point, which constitute the only fermions near the Fermi level within a large 2 eV energy window. Strikingly, the structural chirality rigidly dictates the sign of the topological charges, yielding two ultra-long Fermi arcs spanning the surface Brillouin zone. Our work provides a complete crystallographic classification and a definitive material platform for exploring minimal heterogeneous chiral fermions.

---

*These authors contributed equally to this work.
†Corresponding authors: yangao9419@ysu.edu.cn

Topological semimetals (TSMs) have emerged as a frontier for exploring fundamental physics and developing next-generation quantum devices[1-3]. Depending on the dimensionality of band crossing points in the Brillouin zone (BZ), TSMs are classified into nodal-point, nodal-line, and nodal-surface semimetals[4-7]. Weyl semimetals (WSMs)[8] and Dirac semimetals (DSMs)[9, 10], characterized by zero-dimensional nodal points in momentum space, represent two archetypal classes that have attracted intensive research due to their remarkable properties, including Fermi arc surface states, the chiral anomaly, and giant magnetoresistance[11-15]. Particularly within the study of chiral nodal-point semimetals (CNSMs) hosting chiral fermions[16, 17], a central pursuit is the realization of the "minimal" configuration[18-21]. Because nodes such as Weyl points (WPs) inherently carry non-zero quantized topological charges ($C$), the Nielsen-Ninomiya no-go theorem strictly prohibits them from existing in isolation; instead, they must emerge in pairs of opposite chirality to ensure exact charge neutrality within the BZ[22, 23]. Consequently, the absolute minimal limit of a CNSM is a system hosting exactly one single pair of such nodes—traditionally realized as a single pair of WPs[24-27]. This pristine configuration is highly coveted, as it precludes the complications of multiple nodes, thereby providing the cleanest possible arena to isolate and explore fundamental topological phenomena.

To date, the realization of single-pair CNSMs has been overwhelmingly confined to these homogeneous Weyl systems[24-27]. However, isolated Dirac points (DPs) can also carry non-zero topological charges when protected by specific crystalline symmetries[28, 29]. This raises a natural opportunity to pair a DP with a WP. Consequently, this leads to a critical, unresolved question: Is it possible to realize an intrinsic TSM composed of a single heterogeneous pair—specifically, one WP and one DP? Constructing such a single-pair Weyl-Dirac semimetal imposes a severe crystallographic constraint. To satisfy the no-go theorem, the accompanying DP must carry an equal and opposite topological charge to exactly cancel the intrinsic chiral charge of the WP within the BZ.

Based on the encyclopedia of emergent particles proposed by Yu *et al.*[29], DPs carrying non-zero topological charges across the 230 space groups are strictly restricted to magnitudes of $\mid C \mid = 2$ or $\mid C \mid = 4$. Crucially, a $\mid C \mid = 4$ DP can only emerge in a spinful system (with spin-orbit coupling (SOC))[29]. Conversely, a compatible $\mid C \mid = 4$ WP is strictly confined to spinless systems (without SOC)[29]. This dichotomy creates a mutually exclusive physical condition, strictly forbidding their simultaneous coexistence in a nonmagnetic crystal. Consequently, the only mathematically viable combination for a minimal heterogeneous pair is a $\mid C \mid = 2$ WP coexisting with a $\mid C \mid = 2$ DP. To date, however, the realization of such a single-pair Weyl-Dirac semimetal has remained elusive in both theory and experiment for electronic materials. More importantly, a systematic theoretical framework capable of constructing and predicting this exotic single-pair $\mid C \mid = 2$ Weyl-Dirac coexistence in all 1651 magnetic space groups (MSGs) is fundamentally lacking.

Here, we resolve this fundamental challenge. By systematically imposing the exact little-group constraints for charge-2 Weyl and Dirac points at high-symmetry points and lines based on the encyclopedia of emergent particles across all 1651 MSGs, we reveal that only 14 MSGs without SOC and 10 MSGs with SOC are compatible with this exotic state. For nonmagnetic crystals, this universal theoretical framework identifies the spinless limit of chiral space groups 92 and 96 as the unique hosts for this minimal heterogeneous state. Guided by this framework, we identify enantiomorphic three-dimensional (3D) boron allotropes (SDHBN-$B_{28}$) as the ideal material realization. First-principles calculations unveil an exceptionally clean electronic structure: a charge-2 WP at the $\Gamma$ point and a charge-2 DP at the A point constitute the exclusive band crossings within a massive 2.0 eV energy window around the Fermi level. Strikingly, the structural chirality rigidly dictates the sign of the topological charges, establishing a direct correspondence between real-space handedness and momentum-space topology. The nontrivial topology manifests through remarkable double-helical Fermi arcs that extend across the surface BZ, providing unambiguous experimental signatures. Our results not only demonstrate the first electronic realization of a single-pair charge-2 Weyl–Dirac semimetal but also establish a complete theoretical paradigm for designing minimal heterogeneous chiral fermions.

**Results**

**Symmetry criteria for a single-pair charge-2 Weyl-Dirac composite.** To systematically construct this minimal single-pair configuration, we establish the exact crystallographic prerequisites. Assuming that a $|C| = 2$ WP and a $|C| = 2$ DP are located at momenta $k_W$ and $k_D$ with their corresponding little groups denoted as $G_W$ and $G_D$, the host magnetic space group (MSG) $G$ must satisfy two rigid constraints:

(i) The MSG $G$ must inherently allow for the coexistence of both a $|C| = 2$ WP and a $|C| = 2$ DP.

(ii) To ensure these nodes form an isolated single pair without being multiplied by crystal symmetries, the momenta $k_W$ and $k_D$ must be invariant under all symmetry operations of $G$. This requires their respective little groups to be identical to the group $G$, yielding the exact condition $G_W = G_D = G$.

By exhaustively imposing these little-group constraints onto the encyclopedia of emergent particles across all 1651 MSGs[29-31], we perform a complete theoretical screening for an isolated single-pair composite in regimes both with and without SOC. We reveal that only 14 MSGs without SOC and 10 MSGs with SOC are compatible with this exotic single-pair state. This comprehensive classification is cataloged in Table I, which details the candidate MSGs, the high-symmetry points or lines (HSPs/HSLs) harboring the nodes, and their specific irreducible (co)representations (IRRs). Remarkably, when restricting our search to nonmagnetic crystals (Type II MSGs), the constraints become exceptionally rigid. We rigorously prove that an isolated charge-2 Weyl-Dirac pair can exclusively manifest in the spinless regime within the enantiomorphic chiral SG 92 and 96. Under these unique symmetries, the $|C| = 2$ WP is strictly pinned at the BZ center ($\Gamma$ point), and the $|C| = 2$ DP is enforced at the BZ boundary (A point).

**Material realization in chiral boron networks.** To physically realize this pristine minimal configuration, we turn to elemental boron. Boron exhibits rich structural diversity and, owing to its light atomic mass, possesses negligible SOC, providing an ideal platform to explore the spinless regime[32-36]. Guided by the above symmetry constraints, we designed two enantiomeric boron allotropes that transform into each other under spatial inversion [Figs. 1(a) and 1(b)]. These structures, designated as SDHBN-$B_{28}$ (Single- and Double-atom-wide Helical Boron Network), contain 28 boron atoms per primitive cell [Fig. 1(g)] and exhibit an intricate 3D covalent network based on interwoven helical motifs. The crystal lattice crystallizes in a chiral tetragonal system, where the left-handed enantiomer ($l$-SDHBN-$B_{28}$) [Fig. 1(a)] belongs to SG $P4_32_12$ (No. 96) and its right-handed counterpart ($r$-SDHBN-$B_{28}$) [Fig. 1(b)] belongs to SG $P4_12_12$ (No. 92). The fundamental architecture of both structures consists of an alternating arrangement of single-atom-wide [Figs. 1(d,e)] and double-atom-wide [Figs. 1(c,f)] helical boron chains that wind along the $c$-axis. The optimized lattice parameters for both enantiomers are $a = b = 6.66$ Å and $c = 6.00$ Å, with boron atoms occupying four distinct Wyckoff positions [see Table SI in the Supplemental Material (SM)].

To verify structural stability, we performed phonon dispersion and elastic-constant calculations. Both $l$-SDHBN-$B_{28}$ and $r$-SDHBN-$B_{28}$ exhibit no imaginary phonon frequencies throughout the BZ [Fig. S1 in the SM], confirming their dynamical stability. In addition, the calculated elastic stiffness tensors of the two enantiomers satisfy all Born stability criteria for tetragonal crystals [Table SII in the SM][37], thereby establishing their mechanical stability. These results demonstrate that the proposed structures are promising candidates for experimental synthesis, despite their complex architecture.

**Electronic structure and topological charge characterization.** We subsequently examine the electronic structure of the left-handed enantiomer, $l$-SDHBN-B$_{28}$, using first-principles calculations. As shown by the projected density of states (PDOS) in Fig. 2(b), the system exhibits definitive semimetallic behavior with a profound minimum near the Fermi level ($E_F$). Remarkably, within a massive 2.0 eV energy window centered around $E_F$, only four bands (bands 41–44) are present. This forms an exceptionally clean and isolated four-band manifold, providing an ideal arena for topological analysis.

Detailed momentum-space scanning reveals exactly two distinct band crossing points located at time-reversal invariant momenta (TRIM). First, bands 42 and 43 cross at the $\Gamma$ point ($E_{WP} = -0.020$ eV) to form a twofold degeneracy [Fig. 2(a)]. Under the $D_4$ point group symmetry, the band degeneracy at $\Gamma$ arises from the two-dimensional irreducible representation $\Gamma_5$. This symmetry protection intrinsically dictates a WP. Second, all four bands (41–44) converge at the A point ($E_{DP} = 0.274$ eV) to create a fourfold degeneracy [see Table SIII in the SM]. At this momentum, the little group also maintains $D_4$ symmetry, and the degeneracy arises from the intersection of two doubly degenerate states with irreducible representations $A_1$ and $A_2$, which form a Kramers pair under time-reversal symmetry.

The momentum-space dispersions further validate the unconventional nature of these nodes. Near the $\Gamma$ point, the band structure in the $k_x$-$k_y$ plane exhibits a characteristic quadratic transverse dispersion [Fig. 2(c)], which sharply contrasts with the linear dispersion of conventional $|C| = 1$ WPs and serves as a fundamental hallmark of charge-2 WPs. Conversely, the dispersion near the DP at A displays linear behavior in all three momentum directions [Figs. 2(a) and 2(d)]. To rigorously quantify their topological nature, we tracked the evolution of Wannier charge centers (WCCs) on closed spherical surfaces enclosing each node. For the WP at $\Gamma$, the WCCs complete exactly two full windings, unambiguously establishing a topological charge of $C_{WP} = +2$ [Fig. 2(e)]. Correspondingly, the DP at A exhibits two reversed windings, yielding $C_{DP} = -2$. The Berry curvature distribution in the $(k_x + k_y)$-$k_z$ plane visually corroborates this configuration, depicting the WP as a distinct source of Berry flux and the DP as a sink [Fig. 2(f)]. Because the energy gap between bands 42 and 43 remains finite throughout the rest of the BZ [Fig. 2(g)], this minimal $|C| = 2$ heterogeneous pair perfectly satisfies the exact topological neutrality mandated by the no-go theorem.

**Chirality-topology correspondence and enantiomer comparison.** A defining hallmark of SDHBN-B$_{28}$ is the deterministic mapping between its structural handedness and it's the sign of the momentum-space topological charge. We analyzed the right-handed enantiomer, $r$-SDHBN-B$_{28}$ (SG 92), which is generated from $l$-SDHBN-B$_{28}$ via spatial inversion ($\mathcal{P}$). While the bulk band dispersions, the PDOS, and the irreducible representations at the nodes remain completely invariant under spatial inversion [Figs. 3(a-d)], the topological charges undergo a rigorous sign reversal. The WP at $\Gamma$ now carries a charge of $C_{\mathrm{WP}} = -2$, whereas the DP at A carries $C_{\mathrm{DP}} = +2$ [Fig. 3(e)].

This sign inversion is governed by fundamental symmetry transformation rules. The topological charge is mathematically defined as the surface integral of the Berry curvature. Under spatial inversion $\mathcal{P}$, the Berry curvature, being an axial pseudovector, is mapped from $\boldsymbol{\Omega}_l(\boldsymbol{k})$ to $\boldsymbol{\Omega}_r(-\boldsymbol{k})$. At the same time, the differential surface element transforms to the oppositely oriented element at the inverted momentum, namely from $d\mathbf{S}_l(\boldsymbol{k})$ to $-d\mathbf{S}_r(-\boldsymbol{k})$. As a consequence of these combined transformations, the topological invariant undergoes a strict sign reversal, such that $C_l = -C_r$. This establishes a pristine one-to-one correspondence between the real-space atomic chirality and the momentum-space topological chirality, opening realistic pathways for designing materials with geometrically engineered topological responses.

**The $\boldsymbol{k}\cdot\boldsymbol{p}$ Effective Models for SDHBN-B$_{28}$.** To elucidate the topological nature of the band crossings in SDHBN-B$_{28}$, we perform a symmetry analysis and construct low-energy $\boldsymbol{k}\cdot\boldsymbol{p}$ Hamiltonians describing the $|C| = 2$ WP and DP features. We first consider the WP located at the $\Gamma$ point of the chiral space group $P4_32_12$ (No. 96), whose little group corresponds to the point group $D_4$. The relevant symmetry generators include the screw rotation $\{C_{4,001}^+ \,|\, \frac{1}{2}\frac{1}{2}\frac{3}{4}\}$, $\{C_{2,100} \,|\, \frac{1}{2}\frac{1}{2}\frac{1}{4}\}$, and time-reversal symmetry $\mathcal{T}$. The WP is formed by the crossing of two bands that are derived from the 2D IRR $\Gamma_5$ of the $D_4$ little group [Fig. 2(a)]. Adopting the basis of the $\Gamma_5$ IRR, the representation matrices for the generating elements are expressed as:

$$D\{C_{4,001}^+ \left| \tfrac{1}{2} \tfrac{1}{2} \tfrac{3}{4}\right\} = \begin{bmatrix} 0 & -1 \\ 1 & 0 \end{bmatrix},$$

$$D\{C_{2,100} \left| \tfrac{1}{2} \tfrac{1}{2} \tfrac{1}{4}\right\} = \begin{bmatrix} 0 & -1 \\ -1 & 0 \end{bmatrix},$$

$$D(\mathcal{T}) = \sigma_x K, \tag{1}$$

where $\sigma_x$ is a Pauli matrix and $K$ denotes the complex conjugate operator. These symmetry operations impose the constraint

$$D(g)H(\boldsymbol{k})D^{-1}(g) = H(g\boldsymbol{k}), \tag{2}$$

where $g$ represents the generators $\{C_{4,001}^+, C_{2,100}, \mathcal{T}\}$. Under these symmetry constraints, the two-band effective Hamiltonian expanded to second order in momentum takes the form

$$H_{\mathrm{WP}}(\boldsymbol{k}) = \varepsilon(\boldsymbol{k}) + [(\alpha k_+^2 + \beta k_-^2)\sigma_+ + \mathrm{H.\,c.}] + mk_z\sigma_z, \tag{3}$$

where $\varepsilon(\boldsymbol{k}) = \epsilon_0 + \epsilon_1(k_x^2 + k_y^2) + \epsilon_2 k_z^2$, $k_\pm = k_x \pm ik_y$, $\sigma_\pm = (\sigma_x \pm i\sigma_y)/2$, and $\sigma_i\ (i = x, y, z)$ are there Pauli matrices. The parameters $\epsilon_i(i = 0,1,2)$ and $m$ are real, while $\alpha$ and $\beta$ are complex coefficients determined by the microscopic details of the material. Obviously, the Hamiltonian in Eq. (3) exhibits linear dispersion along the $k_z$ direction and quadratic dispersion in the $k_x - k_y$ plane, which is the characteristic signature of a double-Weyl fermion carrying topological charge $|C| = 2$. By fitting the parameters of this model to the DFT bands[38] of SDHBN-B$_{28}$, the resulting dispersion shows excellent agreement with the *ab initio* bands near the WP [see Fig. S2(a) in the SM], confirming the validity of the effective description.

At the point A, the crossing bands originate from the two distinct 2D IRRs ($A_1$ and $A_2$) of the $D_4$ little group. Their combination produces a symmetry-protected fourfold-degenerate DP. The symmetry generators remain the same as those at the $\Gamma$ point. Under the $A_1 A_2$ IRR as the basis, the matrix representations of the generators are

$$D\{C_{4,001}^+ \left| \tfrac{1}{2} \tfrac{1}{2} \tfrac{3}{4} \right.\} = \tfrac{1}{\sqrt{2}} (-\sigma_0 \otimes \tau_z + i\sigma_z \tau_0),$$

$$D\left\{C_{2,100} \left| \tfrac{1}{2} \tfrac{1}{2} \tfrac{1}{4} \right.\right\} = -\tfrac{i}{\sqrt{2}} (\sigma_z \otimes \tau_x + \sigma_0 \tau_y),$$

$$D(\mathcal{T}) = \sigma_x \otimes \tau_0 K, \tag{4}$$

where $\sigma_{x,y,z}$ and $\tau_{x,y,z}$ are Pauli matrices. Applying the symmetry constraint of Eq. (2), the effective Hamiltonian near the A point can be written as

$$H_{\text{DP}}(\boldsymbol{k}) = \alpha\Gamma_{0,0} + \beta k_z \Gamma_{z,z} + \gamma[(k_x + k_y)\Gamma_{z,x} + (k_x - k_y)\Gamma_{0,y}], \tag{5}$$

where $\Gamma_{i,j} = \sigma_i \otimes \tau_j$, and $\alpha$, $\beta$, and $\gamma$ are real parameters. This model is four-fold degenerate at $k_z = 0$ and exhibits linear dispersion $k_z \to 0$ in all directions, matching the DFT bands at A exactly [see Fig. S2(b) in the SM], confirming the model's reliability.

Finally, because the two enantiomorphic structures share the same point group symmetry and IRRs, the effective Hamiltonians derived above remain valid for the right-handed structure $r$-SDHBN-B$_{28}$. The inversion of structural chirality merely reverses the sign of the topological charges, while leaving the symmetry-allowed structure of the low-energy Hamiltonian unchanged.

**Ultralong double-helical Fermi arcs and bulk-boundary correspondence.** The definitive manifestation of the $|C| = 2$ chiral nodes in SDHBN-B$_{28}$ is the emergence of double Fermi arcs dictated by the bulk-boundary correspondence. Because the $|C| = 2$ WP and DP are maximally separated at the BZ center ($\Gamma$) and boundary (A), their surface projections necessitate extraordinarily extended

topological boundary modes. Surface band structures and surface spectral function calculations for the (100), (110), and (001) surfaces of $l$-SDHBN-B$_{28}$ [Fig. 4(a-f)] unveil two prominent, topologically protected state branches that completely traverse the entire surface BZ. At the WP energy ($E = -20$ meV), the constant-energy contours reveal giant double Fermi arcs originating from the projected WP and terminating at the projected DP. Constrained by the minimal single-pair configuration, these arcs lack alternative termination partners within a single BZ. They are thus forced to extend continuously into adjacent zones, forming an infinite chain of macroscopic topological channels [see Figs. 4(b), 4(d), and 4(f)].

Furthermore, because the right-handed structure $r$-SDHBN-B$_{28}$ has an electronic structure nearly identical to that of $l$-SDHBN-B$_{28}$ [see Figs. 2 and 3], differing only by the reversal of the node chiralities, similar surface band structures and Fermi arcs are obtained for its (100), (110), and (001) surfaces [Fig. 4(g–l)], with the constant-energy slice slightly shifted to $E = -17$ meV. These extended double Fermi arcs therefore provide a clear surface signature of the single-pair $|C| = 2$ Weyl–Dirac semimetal state in SDHBN-B$_{28}$. The universality of these ultralong double Fermi arcs across all evaluated crystalline surfaces and enantiomers overcomes the typical bulk-state obscuration that plagues conventional WSMs. These macroscopic boundary modes provide an unambiguous and highly accessible signature for angle-resolved photoemission spectroscopy (ARPES) measurements, cementing SDHBN-B$_{28}$ as a pristine material platform for observing the boundary excitations of minimal heterogeneous chiral fermions.

**Discussion**

The paramount contribution of this work is the establishment of a complete crystallographic classification for single-pair charge-2 Weyl-Dirac nodes. By systematically screening all 1651 magnetic space groups, we provide a definitive symmetry framework that dictates the emergence of this minimal heterogeneous topological state. This rigorous classification serves as a universal theoretical roadmap, applicable not only to electronic fermionic platforms but also to bosonic materials[39-42]. While our present study highlights this exotic phase using a pristine, non-magnetic crystal as a proof-of-concept, the derived framework concurrently provides explicit crystallographic guidelines for discovering analogous rare topological states within magnetic materials.

The realization of a heterogeneous WP-DP minimal pair fundamentally distinguishes this state from previously known topological semimetals. While single-pair WSMs have been proposed in magnetic systems[21, 24-27] breaking time-reversal symmetry $\mathcal{T}$ and in specific non-magnetic 3D chiral boron materials[43], the stable coexistence of a WP and a DP as an isolated pair in TSMs is unprecedented. Historically, WPs and DPs were viewed as belonging to distinct symmetry regimes, with the former strictly requiring $\mathcal{PT}$-breaking and the latter typically requiring $\mathcal{PT}$-preservation. By exploiting chiral non-symmorphic symmetries in the spinless regime, our framework resolves this conventional dichotomy, enabling the stable coexistence of these charge-2 multifold fermions while perfectly satisfying the Nielsen-Ninomiya theorem.

Guided by our exhaustive symmetry analysis, which restricts this unique single-pair charge-2 Weyl-Dirac state exclusively to chiral space groups 92 and 96 within the 230 space groups under spinless conditions, we identify SDHBN-B$_{28}$ as the ideal material realization. Elemental boron, characterized by its rich structural polymorphism and negligible SOC, serves as the ultimate platform to approach this required spinless limit[32-36]. Consequently, this metastable chiral boron network provides an exceptionally clean electronic environment. It acts as the perfect structural paradigm for isolating and studying the transport properties of these unconventional fermions, entirely free from the interference of incidental band crossings.

Experimentally, the maximal momentum-space separation of the WP and DP generates extraordinary topological surface states. Rather than localized features, the resulting giant double Fermi arcs extensively traverse the surface BZ. Furthermore, the continuous open arcs observed on the (100), (110), and (001) surfaces provide a multi-faceted topological fingerprint that remains robust against variations in surface termination, as long as the bulk WP and DP do not project onto the same surface momentum. Their large momentum-space span may facilitate direct observation by ARPES and could enhance surface-related transport phenomena.

In summary, we have established a complete crystallographic classification for single-pair charge-2 Weyl-Dirac nodes by systematically screening all 1651 MSGs. Our analysis reveals that only 14 MSGs without SOC and 10 with SOC are compatible with this exotic single-pair charge-2 Weyl–Dirac semimetal state.

Furthermore, for nonmagnetic crystals this minimal heterogeneous configuration is uniquely realized in the spinless limit within the chiral space groups 92 and 96, which directly guiding our discovery of the enantiomorphic boron networks ($l/r$-SDHBN-B$_{28}$) as the ideal proof-of-concept material. In this pristine system, a rigid chirality-topology locking mechanism translates real-space structural handedness into deterministic momentum-space the sign of the topological charges, generating ultra-long double Fermi arcs that extensively traverse the surface BZ. Ultimately, the impact of our classification extends far beyond this specific nonmagnetic realization. By providing explicit crystallographic guidelines applicable to both magnetic electronic materials and bosonic materials, this work establishes a definitive paradigm for exploring single-pair $|C| = 2$ Weyl-Dirac chiral quasiparticles.

**Methods**

**Computational methods:** First-principles density-functional theory (DFT) calculations were carried out using the VASP package[44] with the projector augmented-wave (PAW) method[45]. The exchange-correlation interactions were treated within the generalized gradient approximation (GGA) using the Perdew–Burke–Ernzerhof (PBE) functional[46]. The plane-wave cutoff energy was set to 520 eV, and the Brillouin zone (BZ) was sampled using a Γ-centered $6 \times 6 \times 8$ Monkhorst-Pack $k$-point grid[47]. The energy and force convergence criteria were set to $10^{-6}$ eV and $10^{-3}$ eV/Å, respectively. To evaluate structural stability, the elastic stiffness tensor $C_{ij}$ was derived using the energy-strain relationship[48]. Phonon dispersion spectra of SDHBN-B$_{28}$ were calculated via the finite displacement method using the PHONOPY package[49], employing a $2 \times 2 \times 3$ supercell. To explicitly capture the topological signatures of the system, including the topological charges and the ultra-long Fermi arcs, we constructed a tight-binding (TB) Hamiltonian based on maximally localized Wannier functions (MLWFs), as interfaced with the Wannier90 code[50]. The topological surface state spectra and the corresponding Fermi arcs were evaluated using the surface Green's function method via the WannierTools Packages[51].

## Data availability

The data supporting the findings are displayed in the main text and the Supplementary Information. All raw data are available from the corresponding authors upon request.

## Acknowledgements:

We wish to thank Weikang Wu for helpful discussions. This work was supported by the National Natural Science Foundation of China (Grants No. 12304202), Hebei Natural Science Foundation (Grant No. A2023203007), Science Research Project of Hebei Education Department (Grant No. BJK2024085), and Cultivation Project for Basic Research and Innovation of Yanshan University (No. 2022LGZD001).

## Author contributions:

Hui-Jing Zheng proposed the SDHBN-$B_{28}$ structure and performed the corresponding DFT calculations. Ke-Xin Pang carried out the symmetry analysis. Yan Gao conceived the original ideas. Yun-Yun Bai and Yanfeng Ge analyzed the results. Yan Gao writing-reviewing, conceptualization, supervision, and project administration. All authors discussed the results and commented on the manuscript at all stages.

## Competing interests

# Figures and Tables

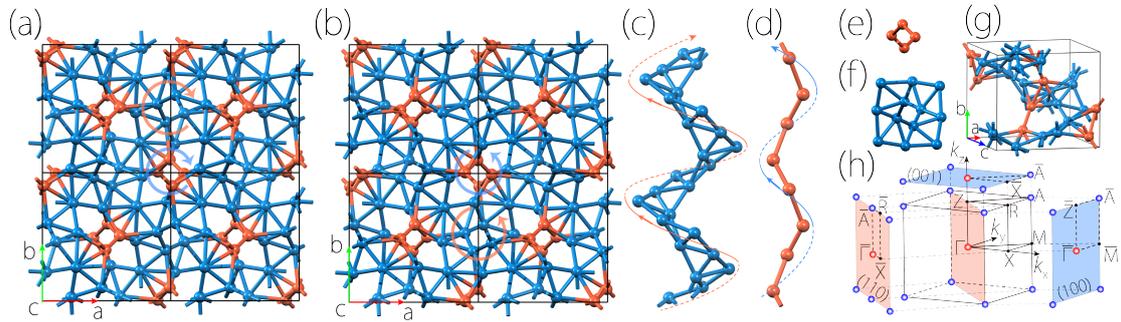

**Figure 1 | Crystal structure and Brillouin zone (BZ) of chiral boron allotrope SDHBN-B$_{28}$.** Perspective views of **a**, left-handed (*l*-SDHBN-B$_{28}$, SG 96) and **b**, right-handed (*r*-SDHBN-B$_{28}$, SG 92) enantiomers. **c**, double helical chain (blue) and **d**, single helical chain (orange) building blocks. Top views of **e**, single and **f**, double helical chains. **g**, Primitive cell showing 28 boron atoms. **h**, Three-dimensional (3D) BZ with charge-2 Weyl point (red) at Γ and charge-2 Dirac point (blue) at A, along with projections onto (001), (100), and (110) surfaces.

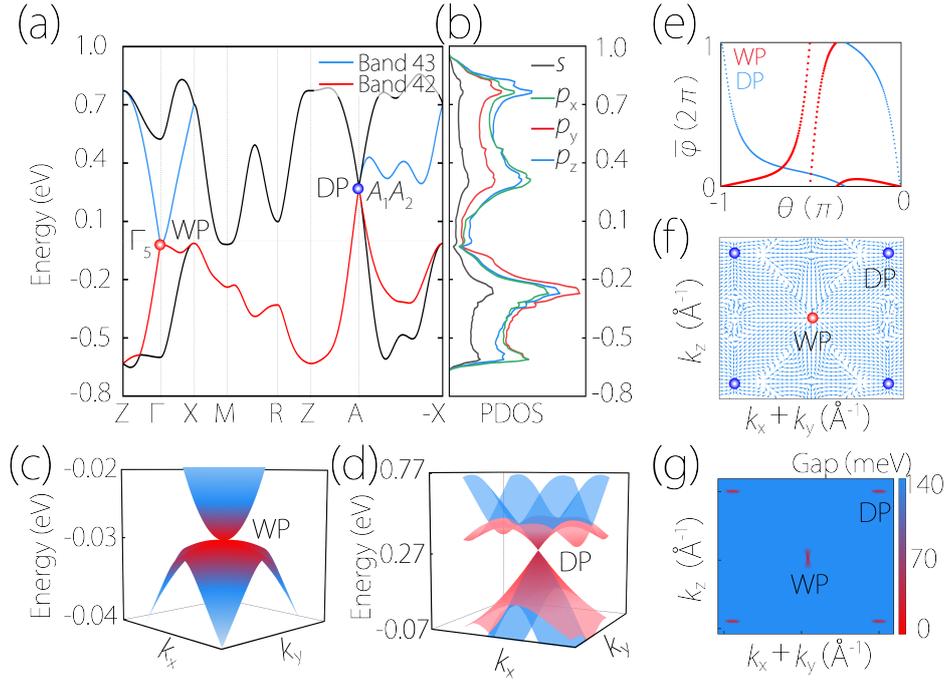

**Figure 2 | Electronic structure and topological characterization of *l*-SDHBN-B$_{28}$. a,** Band structure along high-symmetry lines showing charge-2 Weyl point (red dot, $C = +2$) at $\Gamma$ where bands 42-43 cross, and charge-2 Dirac point (blue dot, $C = -2$) at A where bands 41-44 cross. Irreducible representations (IRRs) are labeled. **b,** Projected density of states (PDOS) showing semimetallic behavior with minimum near Fermi level. 3D band dispersion in $k_x$-$k_y$ plane for **c,** Weyl point (quadratic) and **d,** Dirac point (linear). **e,** Evolution of Wannier charge centers (WCCs) on spheres enclosing Weyl point and Dirac point, revealing topological charges $C = +2$ and $-2$ respectively. **f,** Berry curvature distribution in $(k_x + k_y)$-$k_z$ plane showing Weyl point as source and Dirac point as sink. **g,** Energy gap between bands 42-43 in $(k_x + k_y)$-$k_z$ plane.

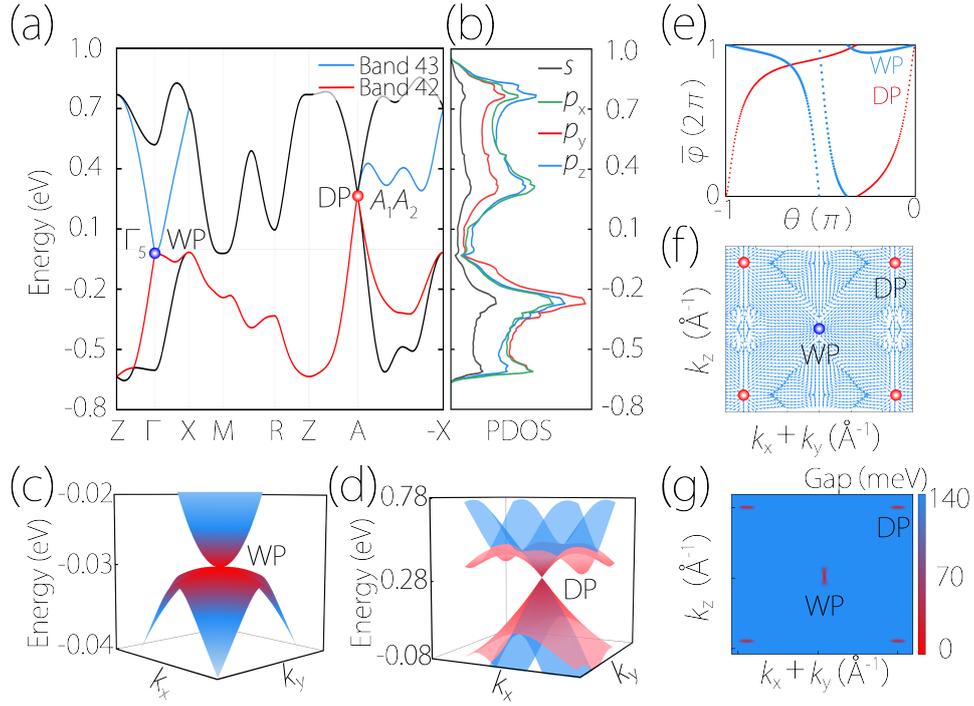

**Figure 3 | Electronic structure and topology characterization of right-handed *r*-SDHBN-B$_{28}$. a**, Band structure showing charge-2 Weyl point (blue dot, $C = -2$) at $\Gamma$ and charge-2 Dirac point (red dot, $C = +2$) at A. **b**, PDOS. 3D band dispersion for **c**, Weyl and **d**, Dirac points. **e**, WCCs evolution revealing reversed topological charges compared to *l*-SDHBN-B$_{28}$. **f**, Berry curvature and **g**, energy gap distribution in $(k_x + k_y)$-$k_z$ plane.

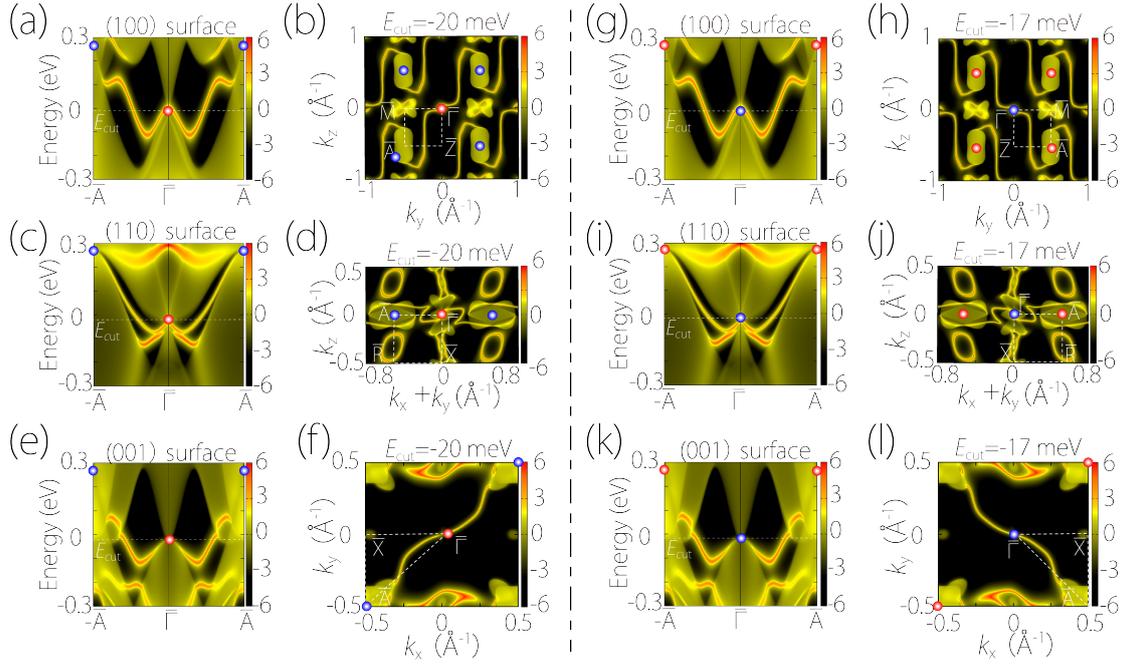

**Figure 4 | Ultralong double-helical Fermi arcs on three surface orientations of *l/r*-SDHBN-B₂₈.** **a-f**, Surface band structures and constant-energy contours (evaluated at an energy slice of $E = -20$ meV) for the *l*-SDHBN-B₂₈ enantiomer. Panels (**a,b**), (**c,d**) and (**e,f**) correspond to the (100), (110) and (001) surfaces, respectively. They reveal continuous, open double Fermi arcs that extensively traverse the entire surface BZ to connect the projected WP and DP, providing a direct visualization of the bulk-boundary correspondence. **g-l**, The corresponding surface band structures and constant-energy contours (evaluated at $E = -17$ meV) for the right-handed *r*-SDHBN-B₂₈ enantiomer across the same three crystal facets.

**TABLE I.** The candidate magnetic space groups (MSGs) that can host a single $|C| = 2$ WP and a single $|C| = 2$ DP. HSP (HSL) denotes the high-symmetry point (line) where the DP or WP is located. IRR stands for the irreducible (co)representation of the little group associated with the respective nodes. The specific momentum-space locations of the single-pair charge-2 Weyl-Dirac nodes under their corresponding host MSGs, together with the generators of these MSGs, are summarized in Table SIV of the Supplemental Material.

| Species | MSGs (HSP or HPL) | Type of MSGs | IRR |
|---|---|---|---|
| | Without SOC | | |
| C-2 DP | 92.112 and 96.144 (A) | II | $\{R_6,R_7\}$ |
| | 90.98, 92.114, 94.130 and 96.146 (MA) | III | $\{V_1V_3,V_2V_4\}$ |
| | 92.115 and 96.147 (A) | III | $T_1T_2$ |
| | 91.109, 91.110, 92.116, 95.141, 95.142 and 96.148 (A) | IV | $A_1A_2$ |
| C-2 WP | 92.112 and 96.114 ($\Gamma$) | II | $R_5$ |
| | 90.98, 92.114, 94.130 and 96.146 ($\Gamma Z$) | III | $\{\Lambda_1,\Lambda_3\}$ or $\{\Lambda_2,\Lambda_4\}$ |
| | 92.115 and 96.147 ($\Gamma$) | III | $\Gamma_2\Gamma_4$ |
| | 91.109, 91.110, 92.116, 95.141, 95.142 and 96.148 ($\Gamma$) | IV | $\Gamma_5$ |
| | With SOC | | |
| C-2 DP | 90.98, 92.114, 94.130 and 96.146 (MA) | III | $\{V_5V_7,V_6V_8\}$ |
| | 92.115 and 96.147 (M) | III | $Y_5Y_5$ |
| | 91.110, 92.116, 95.142 and 96.148 (M) | IV | $M_6M_7$ |
| C-2 WP | 90.98, 92.114, 94.130 and 96.146 ($\Gamma Z$) | III | $\{\Lambda_5,\Lambda_7\}$ or $\{\Lambda_6,\Lambda_8\}$ |
| | 92.115 and 96.147 (Z) | III | $Z_2Z_3$ or $Z_4Z_5$ |
| | 91.110, 92.116, 95.142 and 96.148 (Z) | IV | $Z_7$ |